\documentstyle[epsfig,12pt]{article}

\def\Box{\kern1pt\vbox{\hrule height 1.2pt\hbox{\vrule width 1.2pt\hskip 3pt
   \vbox{\vskip 6pt}\hskip 3pt\vrule width 0.6pt}\hrule height 0.6pt}\kern1pt}
\def\gtwid{\mathrel{\raise.3ex\hbox{$>$\kern-.75em\lower1ex\hbox{$\sim$}}}}
\def\ltwid{\mathrel{\raise.3ex\hbox{$<$\kern-.75em\lower1ex\hbox{$\sim$}}}}
\def\Box{\kern1pt\vbox{\hrule height 1.2pt\hbox{\vrule width 1.2pt\hskip 3pt
   \vbox{\vskip 6pt}\hskip 3pt\vrule width 0.6pt}\hrule height 0.6pt}\kern1pt}
\def\beq{\begin{equation}}
\def\eeq{\end{equation}}
\def\Om{\Omega}
\def\Omp{{\Omega}'}

\def\fio{\varphi_0}

\def\Vp{V_{,\varphi}}

\def\rs{\sqrt{s}}
\def\oots{1-\frac{1}{3s}}
\def\gb{{\widetilde g}}
\def\Hi{H_{\rm i}}

\begin{document}
\begin{titlepage}
\begin{flushright}
astro-ph/9811431 \\ UFIFT-HEP-98-17
\end{flushright}
\vspace{.4cm}

\begin{center}
\textbf{One Loop Back Reaction On Power Law Inflation}
\end{center}

\begin{center}
L. R. Abramo$^{* \dagger}$ and R. P. Woodard$^{\ddagger}$
\end{center}
\begin{center}

\textit{Department of Physics \\ University of Florida \\ 
Gainesville, FL 32611 USA}
\end{center}

\begin{center}
ABSTRACT
\end{center}
\hspace*{.5cm} We consider quantum mechanical corrections to a
homogeneous, isotropic and spatially flat geometry whose scale factor
expands classically as a general power of the co-moving time. The effects
of both gravitons and the scalar inflaton are computed at one loop using
the manifestly causal formalism of Schwinger \cite{Schw} with the Feynman
rules recently developed by Iliopoulos {\it et al.} \cite{Ilio}. We find
no significant effect, in marked contrast with the result obtained by
Mukhanov {\it et al.} \cite{Mukh,Abra} for chaotic inflation based on a
quadratic potential. By applying the canonical technique of Mukhanov {\it
et al.} to the exponential potentials of power law inflation, we show that
the two methods produce the same results, within the approximations
employed, for these backgrounds. We
therefore conclude that the shape of the inflaton potential can have an
enormous impact on the one loop back-reaction. 

\begin{flushleft}
\end{flushleft}
\vspace{.4cm}
\begin{flushleft}
$^*$ Address after May 1, 1999: Theoretische Physik, Ludwig Maximilians \\
Universit\"{a}t, Theresienstr. 37, D-80333 M\"{U}NCHEN, GERMANY \\
$^{\dagger}$ e-mail: abramo@phys.ufl.edu \\
$^{\ddagger}$ e-mail: woodard@phys.ufl.edu
\end{flushleft}
\end{titlepage}

\section{Introduction}

In recent papers, Mukhanov, Abramo and Brandenberger \cite{Mukh,Abra}
have investigated the problem of the back reaction of quantum
fluctuations in chaotic inflation. Using canonical quantization, they
calculated the one loop effective energy-momentum tensor induced by
quantum fluctuations of the matter and metric fields. They found that
the expansion rate of the background Friedmann-Robertson-Walker
space-time slows down as a result of quantum deformations to the
equations of motions.

Is this result generic for all inflationary models? In other words,
does back reaction slow down the expansion rate $H$ in scenarios other
than chaotic inflation? In this paper we extend the work of Mukhanov
{\it et al.} to a very different class of inflationary models, the
so-called power-law inflation \cite{power} in which the expansion of
the universe is given by a scale factor $a(t)\propto t^s$ with $s>1$. 
For that purpose we employ the standard formalism of covariant
quantization, using the Feynman rules for these backgrounds that have
been worked out by Iliopoulos, Tomaras, Tsamis and Woodard \cite{Ilio}.

We find that the one loop back reaction in power-law models is
negligible and stays negligible through the inflation of the universe.
By comparison, in the chaotic models the one loop infrared corrections 
can become non-perturbatively large at late times.

The effect discussed by Mukhanov {\it et al.} is closely related to the
one found in earlier work by Tsamis and Woodard \cite{Tsam} in the
context of pure gravity with a cosmological constant. Both mechanisms
have a simple physical origin: the enhancement of zero-point energy of
the quantum fields by the expansion of the universe, also known in the
literature as {\it superadiabatic amplification}. Virtual quanta of
cosmological wavelengths become trapped by the expansion of the
universe and are unable to recombine. Gravitational interactions
between these virtual pairs, being always attractive, slow the
expansion rate. Because gravitational interactions are very weak, back
reaction takes a long time before it can become a sizeable effect. As
an infrared effect that derives from quantum fluctuations whose
wavelengths are comparable to the Hubble radius $H^{-1}$, back reaction
can be studied in the context of general relativity, whatever is the
ultimate theory of gravity\footnote{This is precisely
the reason why cosmologists are able to predict the seeds of structure
formation from quantum fluctuations in inflationary models without
regard to the true theory of quantum gravity\cite{RevPaper}.}.

A second point of this paper is to prove that results obtained by
covariant quantization should coincide with those worked out using
canonical quantization. This is relevant to the validity and
interpretation of the canonical calculation, which was used
by Mukhanov {\it et al.} to derive their results and has recently been
challenged by Unruh \cite{Unruh}. In a separate paper \cite{Wood} we go
further and show that covariant
quantization yields the same physical results as
canonical quantization in the case of the chaotic inflation models as
well.

The present paper is organized as follows. In section 2 we describe the
perturbative background. We write the Feynman rules for the gravity-scalar
system (taken from Iliopoulos {\it et al} \cite{Ilio}) in section 3. In section
4 we show how to use Schwinger's formalism to obtain expectation values for the
metric and matter fields from the amputated 1-point functions of the theory. In
section 5 we derive the infrared parts of coincident propagators. Our results
are given in section 6. In section 7 we re-derive those results using
gauge-fixed canonical quantization, and show that the two methods give identical
results, namely that one loop back reaction is negligible in power-law
inflation. Section 8 discusses the implications of this work.

\section{Classical background}

The system we consider is general relativity with a
minimally coupled scalar field:

\begin{equation}
\label{Lag}
{\cal L} = {1 \over 16\pi G} \; R \; \sqrt {-g} - \frac12 \; \partial_{\mu} 
\varphi \; \partial_{\nu} \varphi \; g^{\mu \nu} \sqrt {-g} \; - \; V(\varphi) 
\; \sqrt {-g} \; .
\end{equation}
The background is comprised of a flat, homogeneous and isotropic metric

\beq
\label{back_metric}
ds^2_0 = -dt^2 + {a_0}^2 (t) d \vec{x} \cdot d \vec{x}  =
  \Om^2 (\eta) 
  \left( -d\eta^2 + d \vec{x} \cdot d \vec{x} \right)
\eeq
and scalar field $\fio(t)$. The conformal time $\eta$ is related to
comoving time by $dt = \Om (\eta) d \eta$. In this paper we examine
those scalar potentials which drive power-law expansions, that is, the
scale factor grows as a function of time according to

\beq
\label{a_t}
{a_0}(t) = \left( 1+ \frac{\Hi t}{s} \right)^s\; ,
\eeq
where $s$ and $\Hi$ are constants. In terms of conformal time,

\beq
\label{Om_eta}
\Om (\eta) = \left( \frac{\eta_i}{\eta} \right)^{\frac{s}{s-1}} \; ,
\eeq
and $\eta_i = - \frac{s-1}{s} \Hi^{-1}$. The advantage of this class of
expansion laws is that the Feynman rules are already known\cite{Ilio}.

We now write some useful formulas for the background quantities.
The expansion rate is given by the logarithmic derivative of the
scale factor with respect to comoving time,

\beq
\label{exp_rate}
H \equiv \frac{\dot{a}_0}{a_0} = \frac{\Omp}{\Om^2} 
= \Hi \Om^{-\frac{1}{s}} \; ,
\eeq
where a dot indicates $\partial/\partial t$ and
a prime denotes $\partial/\partial\eta={\Om} \partial/\partial t$.

The Einstein field equations are
\begin{eqnarray}  
\label{EFE0}
3 H^2 & = & \frac12 \kappa^2 \left[ \frac12 \dot{\varphi}_0^2 +
V(\varphi_0) \right] \; ,
\\ \label{EFE1}
-2 \dot{H} - 3 H^2 & = & \frac12 \kappa^2 \left[ \frac12 \dot{\varphi}_0^2
- V(\varphi_0)\right] \; ,
\end{eqnarray}
and the integrability condition for this system is given
by the equation of motion for the background scalar field,

\begin{equation}
\label{EOM_f0}
\ddot{\varphi}_0 + 3 H \dot{\varphi}_0 + V_{,\varphi}(\varphi_0) = 0 \; .
\end{equation}
In the expressions above, $\kappa^2=16 \pi G$ is the loop-counting
parameter of perturbative quantum gravity, $V(\varphi)$ is one of the
scalar field potentials of power-law inflation models and 
$\Vp \equiv \partial V/\partial \varphi$.

It is useful to invert the Einstein equations and
write the scalar field quantities in terms of
the Hubble parameter $H$ and its derivative $\dot{H}$:

\begin{eqnarray}
\label{f0d}
\dot{\varphi}_0^2 & = & -{4 \over \kappa^2} \dot{H} \; ,
\\
\label{V_H}
V(\varphi_0) & = & {2 \over \kappa^2} (\dot{H} + 3 H^2) \; .
\end{eqnarray}
In terms of the conformal scale factor, the expressions are

\begin{eqnarray}
\label{f0p}
{\varphi_0'}^2 & = & {4 \over \kappa^2} 
	\left[ - {\Omega^{\prime\prime} \over \Omega} + 
	2 \left( {\Omega^{\prime} \over \Omega} \right)^2 \right] \; ,
\\
\label{V_Om}
V(\varphi_0) & = & {2 \over \kappa^2} {1 \over \Omega^2} 
	\left[ {\Omega^{\prime\prime} \over \Omega} +
	\left({\Omega^{\prime} \over \Omega}\right)^2 \right] \; .
\end{eqnarray}

If we now substitute the expansion law (\ref{Om_eta}) into expressions
(\ref{f0p}) and (\ref{V_Om}), it is easy to obtain the scalar field
potentials which correspond to power-law inflation \cite{power}:

\begin{eqnarray}
\label{f0_Om}
\fio & = & 
	\frac{2}{\rs} \frac{1}{\kappa} \ln{(\Om)} \; , \\
\label{V_fi}
V(\varphi) & = & 
	6 \left( \oots \right) \frac{\Hi^2}{\kappa^2} 
	\exp \left[ -\frac{\kappa}{\rs} \varphi \right] \; .
\end{eqnarray}
The parameter $s$ therefore regulates the steepness of the scalar
potential as well as the rate of expansion. Notice that the equation of
state for the energy density and pressure of the scalar field is

\beq
\label{eq_of_state}
w=\frac{p}{\rho}=\frac{3H^2}{-2\dot{H}-3H^2}=-1+\frac{2}{3s} \; .
\eeq
In the limit $s \rightarrow \infty$ we recover exponential inflation
and the equation of state for de Sitter space, $w = -1$.

\section{Quantum theory: Feynman rules}

Before laying down the quantum theory, we would like to address an
objection that is too often raised: that general relativity is not a
perturbatively consistent quantum theory of gravity, and so cannot be
used to study quantum effects.

That judgment is only partially correct. The back reaction of quantum
fluctuations described here is an {\it infrared} effect, caused by
fluctuations with cosmological wavelengths, and we know, for example
from the Fermi theory of neutrinos\cite{Feinberg}, 
that infrared physics {\it can} be
studied by the low-energy effective theory, regardless of the
renormalizability of that theory.  Conversely, whatever the ultraviolet
behavior of the true quantum theory of gravity may turn out to be, it
will not be affected by these infrared effects. As long as we are careful
not to introduce a spurious time dependence through the ultraviolet
regularization, back reaction is given at late times by ultraviolet
finite terms whose form is entirely controlled by the low-energy theory.

We also note that quantizing some of the gravitational degrees of freedom 
is crucial to {\it all} derivations of the density perturbations
and cosmic microwave background anisotropies in
inflationary cosmology\cite{RevPaper}. Both we and Mukhanov {\it et
al.} just
carry the calculations out to the next stage, and ask what effects quantum 
fluctuations might have on the background in which they propagate.

The goal of this section is to summarize the Feynman rules
for the Lagrangian (\ref{Lag}). The fundamental fields are
the metric and the scalar field,

\begin{eqnarray}
\label{metric}
g_{\mu\nu} & = & \Om^2 ( \eta_{\mu\nu} + \kappa \psi_{\mu\nu} ) 
\equiv \Om^2 \tilde{g}_{\mu\nu} \; , \\
\label{sc_field}
\varphi & = & \varphi_0 + \phi \; .
\end{eqnarray}
Here $\psi_{\mu\nu}$ is the pseudo-graviton field, whose indices are
raised and lowered with the Lorentz metric $\eta_{\mu\nu}$.

Both the scalar-scalar and the graviton-scalar interaction parts of the
Lagrangian can be easily expanded in perturbation theory from the
fundamental Lagrangian (\ref{Lag}). The pure gravitational interactions
are more complicated, however, after some integrations by
parts the whole Lagrangian reduces to the simple expression,

\begin{eqnarray}
\nonumber
{\cal L}_{{\rm inv}} &=&  \Om^2  \sqrt{-\gb}
	\tilde{g}^{\alpha\beta} \gb^{\rho\sigma} \gb^{\mu\nu}
\left[	\frac{1}{2} \psi_{\alpha\rho,\mu} \psi^{\nu\sigma,\beta}
    -	\frac{1}{2} \psi_{\alpha\beta,\rho} \psi^{\sigma\mu,\nu}
    +	\frac{1}{4} \psi_{\alpha\beta,\rho} \psi^{\mu\nu,\sigma} 
\right.
\\
\nonumber
&&- \left.
\frac{1}{4} \psi_{\alpha\rho,\mu}\psi^{\beta\sigma,\nu} \right]	
+ \Om \Omp \sqrt{-\gb} \gb^{\rho\sigma} \gb^{\mu\nu} 
	\psi_{\rho\sigma,\mu} \psi_{0\nu} 
- \Omega^2 \sqrt{-{\gb}} \varphi_0^{\prime} \phi_{,\mu} {\gb}^{0\mu}
\\ 
\label{Lag_g}
&& -  \frac12 \Omega^2  \sqrt{-{\gb}} \phi_{,\mu} \phi_{,\nu} 
	{\widetilde g}^{\mu\nu} 
- \Omega^4 \sqrt{-{\widetilde g}} \sum_{n=1}^{\infty} {1 \over n!} 
{\partial^n V(\varphi_0) \over \partial \varphi^n}  \phi^n \; ,
\end{eqnarray}
up to total derivative terms.

Gauge fixing is achieved by adding a gauge fixing term and the
corresponding ghost action to the invariant Lagrangian:

\begin{equation}
\label{Lag_BRS}
{\cal L}_{\rm BRS} = {\cal L}_{\rm inv} - \frac12 \eta^{\mu \nu} F_{\mu}
F_{\nu} - \Omega \; {\overline \omega}^{\mu} {\delta F}_{\mu} \; .
\end{equation}
The symbol ${\delta F}_{\mu}$ represents the variation of the gauge fixing
functional under an infinitesimal diffeomorphism parameterized by the
ghost field $\omega^{\mu}$. We follow Iliopoulos {\it et al.} in the 
choice of the gauge fixing functional:

\begin{equation}
F_{\mu} = \Omega \left(\psi_{\mu~ ,\nu}^{~\nu} - \frac12 \psi_{,\mu} - 2
{\Omega^{\prime} \over \Omega} \psi_{\mu 0} + \eta_{\mu 0} \kappa
{\varphi}_0^{\prime} \phi \right) \; .
\end{equation}
A great advantage of this gauge is that it decouples the tensor structure
of the propagators from their dependence on spacetime.

The quadratic part of the gauge-fixed Lagrangian 
(\ref{Lag_BRS}) can be written, up to ghost contributions, in terms of a
kinetic operator in ``super-matrix" representation: 

\beq
\label{Def_D}
{\cal D} \equiv 
\left( \begin{array}{ll}
	D_{\mu\nu}^{\;\;\;\;\rho\sigma} & 
		-\frac{2\rs}{s-1} \frac{\Om^2}{\eta^2} t^\rho t^\sigma 
\\
	-\frac{2\rs}{s-1} \frac{\Om^2}{\eta^2} t_\mu t_\nu &
		\Om \left[ \partial^2 + 
		  \frac{2s^2-3s+2}{(s-1)^2} \frac{1}{\eta^2} \right] \Om
	\end{array} \right)  \; ,
\eeq
where $t_\mu=\eta_{\mu 0}$. To get back the quadratic terms of
the Lagrangian (\ref{Lag_BRS}), just multiply the
super-matrix ${\cal D}$ on the left by
$(\psi^{\mu\nu} \; \; \phi)$, and on the right by its transpose.
The kinetic operator $D_{\mu\nu}^{\;\;\;\;\rho\sigma}$ is
given by

\begin{eqnarray}
\nonumber
D_{\mu\nu}^{\;\;\;\;\rho\sigma} & \equiv & 
	\left[ \frac12 \bar\delta_\mu^{(\rho} \bar\delta_\nu^{\sigma)}
	- \frac14 \eta_{\mu\nu} \eta^{\rho\sigma}
	- \frac12 t_\mu t_\nu t^\rho t^\sigma 
	\right] D_A \\
\label{D_munu}
	& & 
	+ \left[ t_\mu t_\nu t^\rho t^\sigma
	- t_{(\mu} \bar\delta_{\nu)}^{\; (\rho} t^{\sigma )}
	\right] D_B \; ,
\end{eqnarray}
where barred tensor symbols denotes that the zero components have been
projected out: $\bar\delta_{\mu}^{\nu} \equiv \delta_{\mu}^{\nu} +
t_\mu t^\nu$. The kinetic operators above are given by

\begin{eqnarray}
\label{D_A}
  D_A^s & \equiv & \Om \left[ 
  \partial^2 + \frac{2s^2-s}{(1-s)^2} \frac{1}{\eta^2} \right] \Om \; , \\
\label{D_B}
  D_B^s & \equiv & \Om \left[ 
  \partial^2 + \frac{s}{(1-s)^2} \frac{1}{\eta^2} \right] \Om \; .
\end{eqnarray}

The off-diagonal term in (\ref{Def_D}), coupling $\psi_{00}$ to $\phi$,
can be removed by a change of variables \cite{Ilio}:

\begin{eqnarray}
\label{zeta}
	\zeta_{ij} & \equiv & \psi_{ij} - \delta_{ij} \psi_{00} \; , \\
\label{chi}
	\chi & \equiv & \phi \sin \theta \ + \psi_{00} \cos \theta \; , \\
\label{upsilon}
	\upsilon & \equiv & \phi \cos \theta - \psi_{00} \sin \theta \; ,
\end{eqnarray}
where $\tan^2 \theta = s$ and Latin letters denote spatial indices.

In terms of the new fields (\ref{zeta})-(\ref{upsilon}) the quadratic,
gauge-fixed Lagrangian is:

\begin{eqnarray}
\nonumber
{\cal L}_{BRS}^{(2)} & = & \frac{1}{2} \zeta_{ij} D_A^s \zeta_{rs} 
\left[ 	\frac{1}{2} \delta_{i(r} \delta_{s)j} -
	\frac{1}{4} \delta_{ij} \delta_{rs}  \right] +
	\frac{1}{2} \zeta_{0i} D_B^s \zeta_{0j} 
\left[	-\delta_{ij} \right] \\
\label{Diag_Lag}
&& 	+ \, \frac{1}{2} \upsilon D_C^s \upsilon
	+ \frac{1}{2} \chi D_A^s \chi 
	+ {\overline \omega}^{i} D_A^s \omega_i
	+ {\overline \omega}^{0} D_B^s \omega_0
\; ,
\end{eqnarray}
where $D_A^s$ and $D_B^s$ were defined above in
(\ref{D_A})-(\ref{D_B}), and $D_C^s$ is given by

\beq
\label{D_C}
  D_C^s \equiv \Om \left[ 
  \partial^2 + \frac{2-s}{(1-s)^2} \frac{1}{\eta^2} \right] \Om \; .
\eeq

The quantization of this system is trivial. There are three modes
associated with the three kinetic operators:

\beq
\label{D_Psi}
D_I^s \Psi (\eta,k,I) = 0 \; .
\eeq
We note for now that by Eq.  (\ref{D_Psi}) and the definitions of the kinetic
operators (\ref{D_A})-(\ref{D_B}) and (\ref{D_C}), the mode functions are
proportional to Hankel functions $H_{\nu_I}^{(1,2)} (k\eta)$ where the labels
$\nu_I(s)$ depends on the parameter $s$. In the limit where $k\eta \rightarrow
\infty$ we find that the modes proportional to $H^{(2)}_{\nu_I}$ have negative
frequency, while those proportional to $H^{(1)}_{\nu_I}=H^{(2)\ast}_{\nu_I}$
have positive frequency.

The fundamental fields $\psi$ and $\phi$ can be expressed in terms
of the mode functions above. The scalar field, for example, is decomposed
in the following manner:

\begin{eqnarray}
\label{phi_mod}
\phi(\eta,\vec{x})
&=& \sum_{I} f_I \psi_I (\eta,\vec{x}) 
\\ \nonumber
&=& \sum_{I} f_I \int \frac{d^3k}{(2\pi)^3}
	\left[ 
	e^{i\vec{k}\cdot\vec{x}}
	\Psi(\eta,k,I) a(\vec{k},I) + 
	e^{-i\vec{k}\cdot\vec{x}}
	\Psi^\ast (\eta,k,I) a^\dagger(\vec{k},I) \right] \; .
\end{eqnarray}
The weights $f_I$ can be found by inverting relations
(\ref{chi})-(\ref{upsilon}), so
in the example above their values are $f_A = \sin{\theta}$, $f_C =
\cos{\theta}$ and $f_B=0$.

The quanta of negative energy are associated with
the negative frequency modes $\Psi$ and the annihilation
operators $a(\vec{k},I)$.
Therefore the vacuum of the theory $|0\rangle$ is defined as the state
that is annihilated by all $a(\vec{k},I)$'s and $I=A,B,C$.

The creation and annihilation operators $a^\dagger$ and $a$ obey the
usual commutation relations

\beq
\label{can_a}
\left[ a(\vec{k},I),a^\dagger(\vec{k}',I') \right] =
	\delta_{II'} \delta^{(3)} (\vec{k}-\vec{k}') \; ,
\eeq
and the mode functions are normalized by

\beq
\label{norm}
\Psi(\eta,k,I) {\Psi '}^{\ast} (\eta,k,I) - 
\Psi^\ast (\eta,k,I) \Psi'(\eta,k,I) = i \Om^{-2} \; .
\eeq

The scalar field propagator is defined by

\beq
\label{Sc_prop}
i\Delta_\phi (x;x') \equiv 
\langle 0 | T \left\{ \phi(x) \phi(x') \right\} |0\rangle
\; .
\eeq
We shall need pure graviton and mixed propagators as well.
All these can be expressed in terms of propagators of
the diagonal fields $\langle 0| T\{\psi_I(x) \psi_I(x') \} |0\rangle$,
which can be conveniently written in momentum space as

\begin{eqnarray}
\label{Prop_I}
i\Delta_I (x;x') = \int \frac{d^3k}{(2\pi)^3} 
	e^{-\varepsilon k} e^{i \vec{k} \cdot (\vec{x}-\vec{x}')} 
	& \times & 
	\left[ 
	\theta(\eta-\eta') \Psi(\eta,k,I) \Psi^\ast(\eta',k,I)
	\right.
	\\ \nonumber
	& & +
\left. \theta(\eta'-\eta) \Psi^\ast(\eta,k,I) \Psi(\eta',k,I) \right]
\; ,
\end{eqnarray}
with $I=A,B,C$. Explicit formulae for these propagators in the infrared
limit can be found in section 5.

Finally, we transform from the diagonal variables $\zeta_{ij}$, $\upsilon$
and $\chi$ back to the original pseudo-graviton $\psi_{\mu\nu}$ and scalar
field $\phi$ by using relations (\ref{zeta})-(\ref{upsilon}). The set of
propagators below is the main result of this section:

\begin{eqnarray}
\nonumber
\langle 0| T \left\{ \psi_{ij}(x)\psi_{rs}(x') \right\} |0\rangle
&=& i \Delta_A^s(x;x') 2 
\left[ \delta_{i(r}\delta_{s)j} - \delta_{ij}\delta_{rs} \right] \\
\nonumber
&& + i \left[ \Delta_A(x;x') + s \Delta_C^s(x;x') \right] 
	\frac{1}{1+s} \delta_{ij}\delta_{rs} \; , \\
\nonumber
\langle 0| T \left\{ \psi_{0i}(x)\psi_{0r}(x') \right\} |0\rangle
&=& - i \Delta_B^s(x;x') \delta_{ir} \; , \\
\nonumber
\langle 0| T \left\{ \psi_{00}(x)\psi_{00}(x') \right\} |0\rangle
&=& i \Delta_A^s(x;x') \frac{1}{1+s} +
  i \Delta_C^s(x;x') \frac{s}{1+s} \; , \\
\nonumber
\langle 0| T \left\{ \psi_{00}(x)\psi_{ij}(x') \right\} |0\rangle
&=& \langle 0| T \left\{ \psi_{00}(x)\psi_{00}(x') \right\} |0\rangle
\delta_{ij} \; , \\
\nonumber
\langle 0| T \left\{ \psi_{00}(x)\phi(x') \right\} |0\rangle
&=& -i \Delta_A^s(x;x') \frac{\rs}{1+s} +
  i \Delta_C^s(x;x') \frac{\rs}{1+s} \; , \\
\nonumber
\langle 0| T \left\{ \psi_{ij}(x)\phi(x') \right\} |0\rangle 
&=& \langle 0| T \left\{ \psi_{00}(x)\phi(x') \right\} |0\rangle
\delta_{ij} \; , \\
\nonumber
\langle 0| T \left\{ \phi(x)\phi(x') \right\} |0\rangle 
&=& i \Delta_A^s(x;x') \frac{s}{1+s} +
  i \Delta_C^s(x;x') \frac{1}{1+s} \; . \\
\nonumber
\langle 0| T \left\{ {\overline \omega}_0 (x) {\overline \omega}_0(x')
\right\} |0\rangle 
&=& - i \Delta_B^s(x;x') \; , \\
\label{Prop_metric}
\langle 0| T \left\{ {\overline \omega}_i (x) {\overline \omega}_j(x')
\right\} |0\rangle 
&=& i \Delta_A^s(x;x') \delta_{ij} \; .
\end{eqnarray}

\section{Attaching external lines}

Back reaction can change the dynamics of the homogeneous and
isotropic background in three ways: through a redefinition of the time
slicing (corrections to $\langle \psi_{00} \rangle$),
through a change in the scale factor (corrections to the trace
$\langle \psi_{ii} \rangle$) and through a shift in the scalar field that
drives
inflation. Because the initial state is homogeneous and isotropic,
the expectation values of the pseudo-graviton and scalar are functions
only of time: 

\begin{eqnarray}
\label{Exp_met}
\langle 0 | \kappa \psi_{\mu\nu}(\eta,\vec{x}) | 0 \rangle 
& = &	A(\eta) \bar{\eta}_{\mu\nu} + C(\eta) t_\mu t_{\nu} \; , \\
\label{Exp_scf}
\langle 0 | \kappa \phi(\eta,\vec{x}) | 0 \rangle & = & D(\eta) 
\; .
\end{eqnarray}
The expectation values of the full metric and
scalar field on the state $|0\rangle$ are then

\begin{eqnarray}
\label{Exp_full_met}
\langle 0| ds^2 |0\rangle & = & \Om^2 \left[ - (1-C) d\eta^2 + 
	(1+A) d\vec{x} \cdot d\vec{x} \right] \\
\nonumber
	&=& -dt^2 + a^2(t)d\vec{x} \cdot d\vec{x} \; , 
\\ \label{Exp_full_scf}
\langle 0| \kappa \varphi |0\rangle & = & \frac{2}{\rs} \ln{(\Om)}
	+ D \; ,
\end{eqnarray}
where $t(\eta)=\int^\eta_{\eta_i} \Om(\eta') d\eta'$ is the background
comoving time and the scale factor in comoving time is given by

\begin{equation}
\label{a}
a^2(t)=a_0^2(t) \left\{ 1 + A[\eta(t)] + H(t) \int_0^t dt' C[\eta(t')] 
\right\} \; .
\end{equation}

We don't measure expectation values directly, but rather
combinations of them that constitute physical observables. The
observables should be independent of gauge-fixing\cite{Annals}. 
In the case of homogeneous and isotropic backgrounds, one such physical
observable is the effective expansion rate $H_{\rm eff}$ defined as the
logarithmic derivative of the scale factor with respect to comoving time:

\begin{eqnarray}
H_{\rm eff} & \equiv & \frac{d \ln{a(t)}}{dt} 
\label{H_eff}
\\ \nonumber
	& = & H(t) \left[ 1 + \frac12 C(t) + 
	\frac{\dot{A}(t)}{2H} + \frac{\dot{H}}{2H} \int_0^t dt' C(t') +
\ldots \right] \; .
\end{eqnarray}

Rather than computing the 1-point functions $A$, $C$ and
$D$ directly, we will compute the amputated 1-point functions
instead, and then attach the external lines to find the 1-point
functions. The amputated 1-point functions are defined as

\begin{eqnarray}
\nonumber
\left( \begin{array}{ll}
	D_{\mu\nu}^{\;\;\;\;\rho\sigma} & 
	  -\frac{2\rs}{s-1} \frac{\Om^2}{\eta^2} t^\rho t^\sigma \\
	-\frac{2\rs}{s-1} \frac{\Om^2}{\eta^2} t_\mu t_\nu &
	  \Om \left[ \partial^2 + 
	    \frac{2s^2-3s+2}{(s-1)^2} \frac{1}{\eta^2} \right] \Om
	\end{array} \right)
&\times&
	\left( \begin{array}{ll}
	\langle 0| \kappa \psi_{\rho\sigma} |0\rangle \\
	\; \; \langle 0| \kappa \phi |0\rangle \end{array} \right) 
\\
\label{Amp_f}
\equiv
\left( \begin{array}{ll}
	\alpha(\eta)\bar\eta_{\mu\nu} + \gamma(\eta) t_\mu t_\nu \\
	\quad \quad \quad \delta (\eta) \end{array} \right) & &\; ,
\end{eqnarray}
where the matrix on the left is the same differential operator defined
above in (\ref{Def_D}). Substituting the expectation values
(\ref{Exp_met}) and (\ref{Exp_scf}) gives

\begin{eqnarray}
\label{f_DF}
\alpha(\eta) & = & - \frac14 D_A (A-C) \; , \\
\label{Def_c_small}
\gamma(\eta) & = & -\frac34 D_A (A-C) + D_B C - 
	\frac{2\rs}{s-1} \frac{\Om^2}{\eta^2} D \; , \\
\label{Def_f_small}
\delta(\eta) & = & \Om \left[
   - \frac{d^2}{d\eta^2} + \frac{2s^2-3s+2}{(s-2)^2} \frac{1}{\eta^2}
	\right] \Om D
   - \frac{2\rs}{s-1} \frac{\Om^2}{\eta^2} C \; .
\end{eqnarray}

Since we are calculating expectation values, we must employ
Schwinger's formalism \cite{Schw} instead of the usual rules for ``in-out"
matrix elements. For the one-loop 1-point functions, the only
difference is that in Schwinger's formalism the external lines are
retarded propagators. In order to enforce this choice of 
external propagator we fix retarded boundary conditions 
such that the 1-point functions $A(\eta)$, $C(\eta)$ and $D(\eta)$,
along with their time derivatives, vanish on the initial surface
$\eta=\eta_i$.

The 1-point functions are obtained by attaching the external legs,
i.e., by inverting the coupled differential equations
(\ref{f_DF})-(\ref{Def_f_small}) with appropriate boundary conditions.
After a change of variables and some algebra, one obtains

\begin{eqnarray}
\label{M}
A &=& 	\frac{1}{D_A} \left[ 
	-4\alpha + \frac{1}{s+1} (3\alpha+\gamma) -\frac{\rs}{s+1} \delta 
\right] \\
\nonumber
& &	\quad \quad +\frac{1}{D_C} \left[ 
	\frac{s}{s+1} (3\alpha+\gamma) + \frac{\rs}{s+1} \delta \right] \; , \\
\label{N}
C &=& 	\frac{1}{D_A} \left[ 
	\frac{1}{s+1} (3\alpha+\gamma) - \frac{\rs}{s+1} \delta \right] \\
\nonumber 
& &	\quad \quad + \frac{1}{D_C} \left[ 
	\frac{s}{s+1} (3\alpha+\gamma) + \frac{\rs}{s+1} \delta \right] \; , \\
\label{U}
D &=& 	
	\frac{1}{D_A} \left[ 
	- \frac{\rs}{s+1} (3\alpha+\gamma)  + \frac{s}{s+1} \delta \right] \\
\nonumber 
& & 	\quad \quad + \frac{1}{D_C} \left[ 
	\frac{\rs}{s+1} (3\alpha+\gamma) + \frac{1}{s+1} \delta \right] \; ,
\end{eqnarray}
where $D_A$ and $D_C$ have been defined in (\ref{D_A}) and (\ref{D_C}). 
Notice that since $\alpha$, $\gamma$ and $\delta$ are functions only of time,
the non-local operators $1/D_I$ in the expressions above denote two time
integrations. The explicit representations for the inverted
propagators are

\begin{eqnarray}
\label{int_DA}
\frac{1}{D_A} f(\eta) &=& - \int_{\eta_i}^\eta d\eta' \Om^{-2}(\eta')
	\int_{\eta_i}^{\eta'} d\eta'' f(\eta'') \; , 
\\ \label{int_DC}
\frac{1}{D_C} f(\eta) &=& - \Om^{-2+\frac2s} (\eta) 
	\int_{\eta_i}^\eta d\eta' \Om^{2-\frac4s}(\eta') 
	\int_{\eta_i}^{\eta'} d\eta'' 
	\Om^{-2+\frac2s} (\eta'') f(\eta'')
\; ,
\end{eqnarray}
where the lower limits of the integrals make clear that

\begin{eqnarray}
\nonumber
A(\eta_i)=C(\eta_i)=D(\eta_i)=0 \; , 
\\
\nonumber
A'(\eta_i)=C'(\eta_i)=D'(\eta_i)=0 \; .
\end{eqnarray}

It is useful to derive explicit formulas for the integrals
(\ref{int_DA})-(\ref{int_DC}) when the integrand has the same 
functional form as the amputated 1-point functions. The fastest-growing
contributions to the amputated 1-point functions have a time dependence
$\propto \Om^{4-2/s+\epsilon}$ (see next section), and for such terms
the integrals (\ref{int_DA})-(\ref{int_DC}) reduce to

\begin{eqnarray}
\label{Inv_DA}
\frac{1}{D_A} \left( \Om^{4 -\frac{2}{s} + \epsilon} \right) & = &
	\frac{s}{\Hi^2 \epsilon (3s-1+s\epsilon)} 
\\ \nonumber
	& \times & \left( 
	- \Om^\epsilon 
	+ \frac{3s-1+s\epsilon}{3s-1} 
	- \frac{s\epsilon}{3s-1} \Om^{-\frac{3s-1}{s}} \right) \; ,
\\
\label{Inv_DC}
\frac{1}{D_C} \left( \Om^{4 -\frac{2}{s} + \epsilon} \right) & = &
	\frac{s^2}{\Hi^2 (s+1+s\epsilon) (2s-2+s\epsilon)}
\\ \nonumber
& & \times
\left( 	- \Om^\epsilon 
	+ \frac{2s-2+s\epsilon}{s-3} \Om^{- \frac{s+1}{s} } 
	- \frac{s+1+s\epsilon}{s-3} \Om^{-2\frac{s-1}{s}} \right) \; .
\end{eqnarray}

\section{Infrared parts of coincident propagators}

In the next section we will sum the diagrams that
contribute to the amputated 1-point functions $\alpha$, $\gamma$ and $\delta$. 
At one loop, the amputated 1-point functions are found by
acting with the 3-point vertex operators on the propagators
$i\Delta_I(x;x')$ and then taking the coincidence limit 
$x \rightarrow x'$. In this section we study the
coincidence limits of the propagators and point out those terms which
can contribute to back reaction.

The three diagonal propagators can be expressed in terms of the mode
functions $\Psi(\eta,k,I)$ which obey equations (\ref{D_Psi}). After a
change of variables

\beq 
\label{chg_var} 
\Psi(\eta,k,I) = (k\eta)^w h(k\eta,I) \quad \quad , 
\quad \quad w=\frac{3s-1}{2(s-1)} \; , 
\eeq 
we can solve Eqs. (\ref{D_Psi}) in terms of Hankel functions
$H_{\nu_I(s)}^{(1,2)} (k\eta)$ of the first and second kind, where the
indices $\nu_I(s)$ are given in terms of the parameter $s$ as \cite{Ilio}

\begin{eqnarray}
\label{nu_A}
\nu_A = \frac32 + \frac{1}{s-1} \; , \\
\label{nu_B}
\nu_B = \frac12 + \frac{1}{s-1} \; , \\
\label{nu_C}
\nu_C = \frac12 - \frac{1}{s-1} \; .
\end{eqnarray}

The normalization of the mode functions is given by (\ref{norm}). The
Hankel functions, on their own account, obey the identities

\begin{eqnarray}
\label{Id1_Hankel}
	H_\nu^{(1)^\ast} (y) & = & H_\nu^{(2)} (y) \; , \\
\label{Id2_Hankel}
H_\nu^{(2)} (y) \frac{d}{dy} H_\nu^{(1)} (y) - 
H_\nu^{(1)} (y) \frac{d}{dy} H_\nu^{(2)} (y) &=& \frac{4i}{\pi} \frac1y
\; ,
\end{eqnarray}
which imply that the two linearly independent, normalized mode functions
are given by

\begin{eqnarray}
\label{norm_Psi_H}
\Psi(\eta,k,I) = 
\frac12 \Omega^{-1} \sqrt{\pi\eta} H_{\nu_I}^{(2)} (k\eta) \; , \\
\Psi^\ast (\eta,k,I) = 
\frac12 \Omega^{-1} \sqrt{\pi\eta} H_{\nu_I}^{(1)} (k\eta) \; .
\end{eqnarray}

The propagators are obtained by substituting the mode functions
above into the definition (\ref{Prop_I}). In the coincidence limit 
$\| \vec{x} - \vec{x}' \| \rightarrow 0$, $\eta-\eta' \rightarrow 0$ we
have, after performing a trivial angular integration,

\beq
\label{prop_H}
i \Delta_I (x) = \frac{1}{8\pi} \frac{|\eta|}{\Om^2}
	\int dk \, k^2 e^{-\epsilon k} 
	H_{\nu_I}^{(1)} (k\eta) H_{\nu_I}^{(2)} (k\eta) \; .
\eeq

Since $H_\nu^{(1,2)}(y) \propto y^{-1/2}$ when $y \rightarrow \infty$,
the coincident propagators diverge as $k^2$ in the ultraviolet. After
these divergences have been regularized at the initial value surface
$\eta=\eta_i$, the counterterms should not affect the time evolution. 
Notice that the details of the regularization procedure and the
ultraviolet behavior of the true, renormalizable quantum theory of
gravity are issues immaterial to the long-range phenomena described
by the effective theory, general relativity.

We are mainly interested in the infrared behavior of the effective
theory, since the physical mechanism behind back reaction is
superadiabatic amplification of quantum fluctuations with physical
wavelengths of the order of or bigger than the Hubble radius $H^{-1}$. We
define the infrared as the scales larger than the Hubble scale,

\beq
\label{infrared}
k_p = \frac{k}{\Om} > H(\eta) = \Hi \Om^{1-1/s} \; .
\eeq

In the far infrared the Hankel functions $H^{(2)}_\nu$ can be
approximated by

\beq
\label{approx_H}
H_{\nu}^{(2)} (k\eta) 
	= \frac{\Gamma(\nu)}{i\pi} 
	\left( \frac{2}{k\eta} \right)^\nu + \cdots \; ,
\eeq
so at coincidence the infrared limit of the propagators (\ref{prop_H})
can be written as

\beq
\label{prop_H_2}
i \Delta_I^{(IR)} (x) = \frac{2^{2\nu_I} \Gamma^2(\nu_I)}{8\pi^3}
	\frac{1}{|\eta|\Om^2} \int^{\Hi \Om^{1-1/s}} 
	dk \, (k\eta)^{2-2\nu_I} e^{-\varepsilon k} \; . 
\eeq
One immediately sees that by Eqs. (\ref{nu_A})-(\ref{nu_C}), for large
$s$ the integral is infrared finite for the $B$ and $C$ modes, but
it is divergent for the $A$ mode.

We can cure this infrared divergence by working on a compact spatial
manifold. The momenta are then bounded from below, and the integrals
above should be mode sums. If we set the size of the compact
manifold to be the $\Hi$, the Hubble radius at the initial time $\eta_i$,
the infrared cut-off is given by $k_0 = \Hi$. In terms of
physical wavelengths, the cut-off is the size of the initially
inflating patch redshifted by the expansion of the universe,
$\lambda^{\rm phys}_0 = 2\pi H_i \Om$.

With the cut-off $k_0=\Hi$ bounding the integral (\ref{prop_H_2}) from
below, the propagator $A$ is given in the infrared limit ($k>H\Om$) by

\beq
\label{Prop_A_IR_int}
i\Delta_A^{(IR)} (\eta) = \frac{\Hi^2}{8\pi^2} (s-1)
\left[ \frac{2 \Gamma \left( \frac32 + \frac{1}{s-1} \right)}{\sqrt{\pi}}
	\left( \frac{s-1}{s} \right)^{\frac{s}{s-1}} \right]^2 
	\left( 1 - \Om^{-2/s} \right) \; ,
\eeq
where the numerical factor between square brackets approaches $1$ in the
limit $s \rightarrow \infty $.

The propagators for the $B(+)$ and $C(-)$ modes in the infrared
limit are best left as integrals,

\beq
\label{Prop_B_IR}
i\Delta_{B,C}^{(IR)} = \frac{\Hi^{\pm \frac{2}{s-1}}}{4\pi^2}
\left[ \frac{\Gamma \left( \frac{1}{2} \pm \frac{1}{s-1}
	\right)}{\sqrt{\pi}} \left( \frac{s-1}{s} \right)^{\pm
	\frac{1}{s-1}} \right]^2 \Om^{-2 \pm \frac{2}{s}} 
	\int_{\Hi}^{\Hi \Om^{1-1/s}} dk k^{1\mp\frac{2}{s-1}} \; ,
\eeq
where it is easy to see that they are dominated by the
ultraviolet. In addition, they have an overall time factor of
$\Om^{-2/s}$ which makes them subdominant when compared with the
constant part of $i\Delta_A^{(IR)}$.

The dominant contribution from the infrared limit of the coincident
propagators is therefore 

\beq
\label{Prop_A_IR}
i\Delta_A^{(IR)} = \frac{\Hi^2}{8\pi^2} (s-1)
\left[ \frac{2 \Gamma \left( \frac32 + \frac{1}{s-1} \right)}{\sqrt{\pi}}
	\left( \frac{s-1}{s} \right)^{\frac{s}{s-1}} \right]^2 \; .
\eeq

\section{Results}

In this section we obtain and discuss the results for the expectation
values of the metric and scalar field in the presence of the quantum
fluctuations $\psi_{\mu\nu}$ and $\phi$.

Our strategy consists of summing up all contributions to the amputated
1-point functions (\ref{Amp_f}) coming from cubic interactions, then
obtaining the expectation values of the metric and scalar field by
(\ref{M})-(\ref{U}).

There are three types of vertices with cubic interactions in
(\ref{Lag_BRS}): pure graviton vertices $\psi^3$ (Table 1),
graviton-ghost vertex $\psi \bar{w} w$ (Table 2) and vertices with one
or more $\phi$'s (Table 3).

\begin{table}

\vbox{\tabskip=0pt \offinterlineskip
\def\tablerule{\noalign{\hrule}}
\halign to460pt {\strut#& \vrule#\tabskip=1em plus2em& 
\hfil#& \vrule#& \hfil#\hfil& \vrule#& \hfil#& \vrule#& \hfil#\hfil& 
\vrule#\tabskip=0pt\cr
\tablerule
\omit&height4pt&\omit&&\omit&&\omit&&\omit&\cr
&&\omit\hidewidth \# &&\omit\hidewidth {\rm Vertex Factor}\hidewidth&& 
\omit\hidewidth \#\hidewidth&& \omit\hidewidth {\rm Vertex Factor}
\hidewidth&\cr
\omit&height4pt&\omit&&\omit&&\omit&&\omit&\cr
\tablerule
\omit&height2pt&\omit&&\omit&&\omit&&\omit&\cr
&& 1 && $\frac12 \kappa \Hi \Omega^{3 -1/s} \eta^{\alpha_1 \beta_1}
\eta^{\alpha_2 \beta_2}  \partial_2^{(\alpha_3}  t^{\beta_3)}$ 
&& 22 && $\frac12 \kappa \Omega^2 \eta^{\alpha_2 (\alpha_3}
\eta^{\beta_3) \beta_2}  \partial_3^{(\alpha_1}  
\partial_1^{\beta_1)}$ &\cr
\omit&height2pt&\omit&&\omit&&\omit&&\omit&\cr
\tablerule
\omit&height2pt&\omit&&\omit&&\omit&&\omit&\cr
&& 2 && $\frac12 \kappa \Hi \Omega^{3-1/s} \eta^{\alpha_2 \beta_2}
\eta^{\alpha_3 
\beta_3}  \partial_3^{(\alpha_1}  t^{\beta_1)}$ 
&& 23 && $\frac12 \kappa \Omega^2 \eta^{\alpha_3 (\alpha_1}  \eta^{\beta_1) 
\beta_3}  \partial_1^{(\alpha_2}  \partial_2^{\beta_2)}$ &\cr
\omit&height2pt&\omit&&\omit&&\omit&&\omit&\cr
\tablerule
\omit&height2pt&\omit&&\omit&&\omit&&\omit&\cr
&& 3 && $\frac12 \kappa \Hi \Omega^{3-1/s} \eta^{\alpha_3 \beta_3}
\eta^{\alpha_1 
\beta_1}  \partial_1^{(\alpha_2}  t^{\beta_2)}$ 
&& 24 && $\frac12 \kappa \Omega^2 \partial_2^{(\alpha_1}  \eta^{\beta_1) 
(\alpha_3}  \partial_3^{\beta_3)}  \eta^{\alpha_2 \beta_2}$ &\cr
\omit&height2pt&\omit&&\omit&&\omit&&\omit&\cr
\tablerule
\omit&height2pt&\omit&&\omit&&\omit&&\omit&\cr
&& 4 && $-\kappa \Hi \Omega^{3-1/s} \eta^{\alpha_1 (\alpha_2}
\eta^{\beta_2) \beta_1}
 \partial_2^{(\alpha_3}  t^{\beta_3)}$ 
&& 25 && $\frac12 \kappa \Omega^2 \partial_3^{(\alpha_2}  \eta^{\beta_2) 
(\alpha_1}  \partial_1^{\beta_1)}  \eta^{\alpha_3 \beta_3}$ &\cr
\omit&height2pt&\omit&&\omit&&\omit&&\omit&\cr
\tablerule
\omit&height2pt&\omit&&\omit&&\omit&&\omit&\cr
&& 5 && $-\kappa \Hi \Omega^{3-1/s} \eta^{\alpha_2 (\alpha_3}
\eta^{\beta_3) \beta_2}
 \partial_3^{(\alpha_1}  t^{\beta_1)}$ 
&& 26 && $\frac12 \kappa \Omega^2 \partial_1^{(\alpha_3}  \eta^{\beta_3) 
(\alpha_2}  \partial_2^{\beta_2)}  \eta^{\alpha_1 \beta_1}$ &\cr
\omit&height2pt&\omit&&\omit&&\omit&&\omit&\cr
\tablerule
\omit&height2pt&\omit&&\omit&&\omit&&\omit&\cr
&& 6 && $-\kappa \Hi \Omega^{3-1/s} \eta^{\alpha_3 (\alpha_1}
\eta^{\beta_1) \beta_3}
 \partial_1^{(\alpha_2}  t^{\beta_2)}$ 
&& 27 && $\frac12 \kappa \Omega^2 \partial_2^{(\alpha_1}  \eta^{\beta_1) 
(\alpha_2}  \partial_3^{\beta_2)}  \eta^{\alpha_3 \beta_3}$ &\cr
\omit&height2pt&\omit&&\omit&&\omit&&\omit&\cr
\tablerule
\omit&height2pt&\omit&&\omit&&\omit&&\omit&\cr
&& 7 && $-\kappa \Hi \Omega^{3-1/s} t^{(\alpha_3}  \eta^{\beta_3)
(\alpha_1}
 \partial_2^{\beta_1)}  \eta^{\alpha_2 \beta_2}$ 
&& 28 && $\frac12 \kappa \Omega^2 \partial_3^{(\alpha_2}  \eta^{\beta_2) 
(\alpha_3}  \partial_1^{\beta_3)}  \eta^{\alpha_1 \beta_1}$ &\cr
\omit&height2pt&\omit&&\omit&&\omit&&\omit&\cr
\tablerule
\omit&height2pt&\omit&&\omit&&\omit&&\omit&\cr
&& 8 && $-\kappa \Hi \Omega^{3-1/s} t^{(\alpha_1}  \eta^{\beta_1)
(\alpha_2} 
 \partial_3^{\beta_2)}  \eta^{\alpha_3 \beta_3}$ 
&& 29 && $\frac12 \kappa \Omega^2 \partial_1^{(\alpha_3}  \eta^{\beta_3) 
(\alpha_1}  \partial_2^{\beta_1)}  \eta^{\alpha_2 \beta_2}$ &\cr
\omit&height2pt&\omit&&\omit&&\omit&&\omit&\cr
\tablerule
\omit&height2pt&\omit&&\omit&&\omit&&\omit&\cr
&& 9 && $-\kappa \Hi \Omega^{3-1/s} t^{(\alpha_2}  \eta^{\beta_2)
(\alpha_3} 
 \partial_1^{\beta_3)}  \eta^{\alpha_1 \beta_1}$ 
&& 30 && $\frac18 \kappa \Omega^2 \eta^{\alpha_1 \beta_1}  \eta^{\alpha_2 
\beta_2}  \eta^{\alpha_3 \beta_3}  \partial_2 \cdot \partial_3$ &\cr
\omit&height2pt&\omit&&\omit&&\omit&&\omit&\cr
\tablerule
\omit&height2pt&\omit&&\omit&&\omit&&\omit&\cr
&& 10 && $\frac14 \kappa \Omega^2 \eta^{\alpha_1 \beta_1}  \partial_3^{
(\alpha_2}  \eta^{\beta_2) (\alpha_3}  \partial_2^{\beta_3)}$ 
&& 31 && $\frac14 \kappa \Omega^2 \eta^{\alpha_1 \beta_1}  \eta^{\alpha_2 
\beta_2}  \eta^{\alpha_3 \beta_3}  \partial_3 \cdot \partial_1$ &\cr
\omit&height2pt&\omit&&\omit&&\omit&&\omit&\cr
\tablerule
\omit&height2pt&\omit&&\omit&&\omit&&\omit&\cr
&& 11 && $\frac14 \kappa \Omega^2 \eta^{\alpha_2 \beta_2}  \partial_1^{
(\alpha_3}  \eta^{\beta_3) (\alpha_1}  \partial_3^{\beta_1)}$ 
&& 32 && $-\frac12 \kappa \Omega^2 \eta^{\alpha_1 (\alpha_2}  \eta^{\beta_2) 
\beta_1}  \eta^{\alpha_3 \beta_3}  \partial_2 \cdot \partial_3$ &\cr
\omit&height2pt&\omit&&\omit&&\omit&&\omit&\cr
\tablerule
\omit&height2pt&\omit&&\omit&&\omit&&\omit&\cr
&& 12 && $\frac14 \kappa \Omega^2 \eta^{\alpha_3 \beta_3}  \partial_2^{
(\alpha_1}  \eta^{\beta_1) (\alpha_2}  \partial_1^{\beta_2)}$ 
&& 33 && $-\frac12 \kappa \Omega^2 \eta^{\alpha_2 (\alpha_3}  \eta^{\beta_3) 
\beta_2}  \eta^{\alpha_1 \beta_1}  \partial_3 \cdot \partial_1$ &\cr
\omit&height2pt&\omit&&\omit&&\omit&&\omit&\cr
\tablerule
\omit&height2pt&\omit&&\omit&&\omit&&\omit&\cr
&& 13 && $-\kappa \Omega^2 \partial_3^{(\alpha_1}  \eta^{\beta_1) (\alpha_2} 
 \eta^{\beta_2) (\alpha_3}  \partial_2^{\beta_3)}$ 
&& 34 && $-\frac12 \kappa \Omega^2 \eta^{\alpha_3 (\alpha_1}  \eta^{\beta_1) 
\beta_3}  \eta^{\alpha_2 \beta_2}  \partial_1 \cdot \partial_2$ &\cr
\omit&height2pt&\omit&&\omit&&\omit&&\omit&\cr
\tablerule
\omit&height2pt&\omit&&\omit&&\omit&&\omit&\cr
&& 14 && $-\kappa \Omega^2 \partial_1^{(\alpha_2}  \eta^{\beta_2) (\alpha_3} 
 \eta^{\beta_3) (\alpha_1}  \partial_3^{\beta_1)}$ 
&& 35 && $-\frac14 \kappa \Omega^2 \partial_2^{(\alpha_1}  \partial_3^{
\beta_1)}  \eta^{\alpha_2 \beta_2}  \eta^{\alpha_3 \beta_3}$ &\cr
\omit&height2pt&\omit&&\omit&&\omit&&\omit&\cr
\tablerule
\omit&height2pt&\omit&&\omit&&\omit&&\omit&\cr
&& 15 && $-\kappa \Omega^2 \partial_2^{(\alpha_3}  \eta^{\beta_3) (\alpha_1} 
 \eta^{\beta_1) (\alpha_2}  \partial_1^{\beta_2)}$ 
&& 36 && $-\frac12 \kappa \Omega^2 \partial_3^{(\alpha_2}  \partial_1^{
\beta_2)}  \eta^{\alpha_3 \beta_3}  \eta^{\alpha_1 \beta_1}$ &\cr
\omit&height2pt&\omit&&\omit&&\omit&&\omit&\cr
\tablerule
\omit&height2pt&\omit&&\omit&&\omit&&\omit&\cr
&& 16 && $-\frac12 \kappa \Omega^2 \partial_3^{(\alpha_2}  \eta^{\beta_2) 
(\alpha_1}  \eta^{\beta_1) (\alpha_3}  \partial_2^{\beta_3)}$ 
&& 37 && $-\frac18 \kappa \Omega^2 \eta^{\alpha_1 \beta_1}  \eta^{\alpha_2 
(\alpha_3}  \eta^{\beta_3) \beta_2}  \partial_2 \cdot \partial_3$ &\cr
\omit&height2pt&\omit&&\omit&&\omit&&\omit&\cr
\tablerule
\omit&height2pt&\omit&&\omit&&\omit&&\omit&\cr
&& 17 && $-\kappa \Omega^2 \partial_1^{(\alpha_3}  \eta^{\beta_3) (\alpha_2} 
 \eta^{\beta_2) (\alpha_1}  \partial_3^{\beta_1)}$ 
&& 38 && $-\frac14 \kappa \Omega^2 \eta^{\alpha_2 \beta_2}  \eta^{\alpha_3 
(\alpha_1}  \eta^{\beta_1) \beta_3}  \partial_3 \cdot \partial_1$ &\cr
\omit&height2pt&\omit&&\omit&&\omit&&\omit&\cr
\tablerule
\omit&height2pt&\omit&&\omit&&\omit&&\omit&\cr
&& 18 && $-\frac14 \kappa \Omega^2 \eta^{\alpha_1 \beta_1}  \eta^{\alpha_2 
\beta_2}  \partial_2^{(\alpha_3}  \partial_3^{\beta_3)}$ 
&& 39 && $\frac12 \kappa \Omega^2 \eta^{\alpha_1) (\alpha_2}  \eta^{\beta_2) 
(\alpha_3}  \eta^{\beta_3) (\beta_1}  \partial_2 \cdot \partial_3$ &\cr
\omit&height2pt&\omit&&\omit&&\omit&&\omit&\cr
\tablerule
\omit&height2pt&\omit&&\omit&&\omit&&\omit&\cr
&& 19 && $-\frac14 \kappa \Omega^2 \eta^{\alpha_2 \beta_2}  \eta^{\alpha_3 
\beta_3}  \partial_3^{(\alpha_1}  \partial_1^{\beta_1)}$ 
&& 40 && $\kappa \Omega^2 \eta^{\alpha_1) (\alpha_2}  \eta^{\beta_2) 
(\alpha_3}  \eta^{\beta_3) (\beta_1}  \partial_3 \cdot \partial_1$ &\cr
\omit&height2pt&\omit&&\omit&&\omit&&\omit&\cr
\tablerule
\omit&height2pt&\omit&&\omit&&\omit&&\omit&\cr
&& 20 && $-\frac14 \kappa \Omega^2 \eta^{\alpha_3 \beta_3}  \eta^{\alpha_1 
\beta_1}  \partial_1^{(\alpha_2}  \partial_2^{\beta_2)}$ 
&& 41 && $\frac14 \kappa \Omega^2 \partial_2^{(\alpha_1}  \partial_3^{
\beta_1)}  \eta^{\alpha_2 (\alpha_3}  \eta^{\beta_3) \beta_2}$ &\cr
\omit&height2pt&\omit&&\omit&&\omit&&\omit&\cr
\tablerule
\omit&height2pt&\omit&&\omit&&\omit&&\omit&\cr
&& 21 && $\frac12 \kappa \Omega^2 \eta^{\alpha_1 (\alpha_2}  \eta^{\beta_2) 
\beta_1}  \partial_2^{(\alpha_3}  \partial_3^{\beta_3)}$
&& 42 && $\frac12 \kappa \Omega ^2 \partial_3^{(\alpha_2}  \partial_1^{
\beta_2)}  \eta^{\alpha_3 (\alpha_1}  \eta^{\beta_1) \beta_3}$ &\cr 
\omit&height2pt&\omit&&\omit&&\omit&&\omit&\cr
\tablerule}}

\caption{Vertex factors contracted into $\psi_{\alpha_1\beta_1}
\psi_{\alpha_2\beta_2} \psi_{\alpha_3\beta_3}$ with \#1 external.}

\end{table}

\begin{table}

\vbox{\tabskip=0pt \offinterlineskip
\def\tablerule{\noalign{\hrule}}
\halign to450pt {\strut#& \vrule#\tabskip=1em plus2em& 
\hfil#& \vrule#& \hfil#\hfil& \vrule#& \hfil#& \vrule#& \hfil#\hfil& 
\vrule#\tabskip=0pt\cr
\tablerule
\omit&height4pt&\omit&&\omit&&\omit&&\omit&\cr
&&\omit\hidewidth \# 
&&\omit\hidewidth {\rm Vertex Factor}\hidewidth&& 
\omit\hidewidth \#\hidewidth&& 
\omit\hidewidth {\rm Vertex Factor}
\hidewidth&\cr
\omit&height4pt&\omit&&\omit&&\omit&&\omit&\cr
\tablerule
\omit&height2pt&\omit&&\omit&&\omit&&\omit&\cr
&& 1 && $- \kappa \Omega^2 \eta^{\alpha_2 (\alpha_1} \eta^{\beta_1) 
\alpha_3} \partial_2 \cdot \partial_3$ 
&& 6 && $\frac12 \kappa \Omega^2 \eta^{\alpha_1 \beta_1} \partial_2^{
\alpha_2} \partial_1^{\alpha_3}$ &\cr
\omit&height2pt&\omit&&\omit&&\omit&&\omit&\cr
\tablerule
\omit&height2pt&\omit&&\omit&&\omit&&\omit&\cr
&& 2 && $- \kappa \Omega^2 \eta^{\alpha_3 (\alpha_1} \partial_2^{\beta_1)} 
\partial_3^{\alpha_2}$ 
&& 7 && $-\kappa \Hi \Omega^3 \eta^{\alpha_1 \beta_1}
\partial_2^{\alpha_2} 
t^{\alpha_3}$ &\cr
\omit&height2pt&\omit&&\omit&&\omit&&\omit&\cr
\tablerule
\omit&height2pt&\omit&&\omit&&\omit&&\omit&\cr
&& 3 && $- \kappa \Omega^2 \eta^{\alpha_2 (\alpha_1} \partial_2^{\beta_1)} 
\partial_1^{\alpha_3}$ 
&& 8 && $-2 \kappa \Hi \Omega^3 \eta^{\alpha_3 (\alpha_1} \partial_3^{
\beta_1)} t^{\alpha_2}$ &\cr
\omit&height2pt&\omit&&\omit&&\omit&&\omit&\cr
\tablerule
\omit&height2pt&\omit&&\omit&&\omit&&\omit&\cr
&& 4 && $2 \kappa \Hi \Omega^3 \eta^{\alpha_2 (\alpha_1} \partial_2^{
\beta_1)} t^{\alpha_3}$ 
&& 9 && $-\kappa \Hi \Omega^3 \eta^{\alpha_1 \beta_1}
\partial_1^{\alpha_3} 
t^{\alpha_2}$ &\cr
\omit&height2pt&\omit&&\omit&&\omit&&\omit&\cr
\tablerule
\omit&height2pt&\omit&&\omit&&\omit&&\omit&\cr
&& 5 && $\kappa \Omega^2 \eta^{\alpha_3 (\alpha_1} \partial_3^{\beta_1)} 
\partial_2^{\alpha_2}$ 
&& 10 && $2 \kappa \Hi^2 \Omega^4 \eta^{\alpha_1 \beta_1} t^{\alpha_2} 
t^{\alpha_3}$ &\cr
\omit&height2pt&\omit&&\omit&&\omit&&\omit&\cr
\tablerule}}

\caption{Vertex factors contracted into $\psi_{\alpha_1\beta_1} 
{\overline \omega}_{\alpha_2} \omega_{\alpha_3}$.}

\end{table}

\begin{table}

\vbox{\tabskip=0pt \offinterlineskip
\def\tablerule{\noalign{\hrule}}
\halign to450pt {\strut#& \vrule#\tabskip=1em plus2em& 
\hfil#& \vrule#& \hfil#\hfil& \vrule#& \hfil#& \vrule#& \hfil#\hfil& 
\vrule#\tabskip=0pt\cr
\tablerule
\omit&height4pt&\omit&&\omit&&\omit&&\omit&\cr
&&\omit\hidewidth \# 
&&\omit\hidewidth {\rm Interaction}\hidewidth&& 
\omit\hidewidth \#\hidewidth&& 
\omit\hidewidth {\rm Interaction}
\hidewidth&\cr
\omit&height4pt&\omit&&\omit&&\omit&&\omit&\cr
\tablerule
\omit&height2pt&\omit&&\omit&&\omit&&\omit&\cr
&& 1 && $\frac1{4\sqrt{s}} \kappa H_i \Omega^{3-\frac1{s}}\phi^{\prime} \psi^2$
&& 6 && $\frac12 \kappa \Omega^2 \phi_{,\rho} \phi_{,\sigma} 
\psi^{\rho\sigma}$ &\cr 
\omit&height2pt&\omit&&\omit&&\omit&&\omit&\cr
\tablerule
\omit&height2pt&\omit&&\omit&&\omit&&\omit&\cr
&& 2 && $-\frac1{2 \sqrt{s}} \kappa H_i \Omega^{3-\frac1{s}} \phi^{\prime} 
\psi^{\rho\sigma} \psi_{\rho\sigma}$
&& 7 && $\frac1{4 \sqrt{s}} (3 - \frac1{s}) \kappa H_i^2 \Omega^{4-\frac2{s}} 
\phi \psi^2$ &\cr
\omit&height2pt&\omit&&\omit&&\omit&&\omit&\cr
\tablerule
\omit&height2pt&\omit&&\omit&&\omit&&\omit&\cr
&& 3 && $-\frac1{\sqrt{s}} \kappa H_i \Omega^{3-\frac1{s}} t_{\rho} \phi_{,
\sigma} \psi^{\rho\sigma} \psi$
&& 8 && $-\frac1{2 \sqrt{s}} (3-\frac1{s}) \kappa H_i^2 \Omega^{4-\frac2{s}}
\phi \psi^{\rho\sigma} \psi_{\rho\sigma}$ &\cr 
\omit&height2pt&\omit&&\omit&&\omit&&\omit&\cr
\tablerule
\omit&height2pt&\omit&&\omit&&\omit&&\omit&\cr
&& 4 && $\frac2{\sqrt{s}} \kappa H_i \Omega^{3-\frac1{s}} t_{\rho} \phi_{,
\sigma} \psi^{\rho\mu} \psi^{\sigma}_{~\mu}$
&& 9 && $-\frac1{2 s} (3 - \frac1{s}) \kappa H_i^2 \Omega^{4-\frac2{s}} \phi^2 
\psi$ &\cr 
\omit&height2pt&\omit&&\omit&&\omit&&\omit&\cr
\tablerule
\omit&height2pt&\omit&&\omit&&\omit&&\omit&\cr
&& 5 && $-\frac14 \kappa \Omega^2 \phi_{,\rho} \phi^{,\rho} \psi$
&& 10 && $\frac1{2 s^{3/2}} (3 - \frac1{s}) \kappa H_i^2 \Omega^{4-\frac2{s}} 
\phi^3$ &\cr 
\omit&height2pt&\omit&&\omit&&\omit&&\omit&\cr
\tablerule
\omit&height2pt&\omit&&\omit&&\omit&&\omit&\cr
&& \omit && \omit
&& 11 && $\frac2{\sqrt{s}} \kappa H_i \Omega^{3-\frac1{s}} t_{\rho} \phi_{,
\sigma} {\overline \omega}^{\rho} \omega^{\sigma}$ &\cr 
\omit&height2pt&\omit&&\omit&&\omit&&\omit&\cr
\tablerule}}

\caption{Cubic interactions involving $\phi$.}

\end{table}

For simplicity, we have partially symmetrized the $\psi^3$ vertices of
Table 1 so that the first field always corresponds to the external
line. For example, the following vertex 

\beq
\label{vertex_1}
\frac12 \kappa \Hi \Om^{3-1/s} \psi \psi_{,\alpha} \psi^{\alpha 0} \; ,
\eeq
gives rise to the symmetrized vertices numbers $1$, 2 or 3 of Table 1
when the first, third or second $\psi$'s respectively are taken as the
external leg.

As an example, consider the contributions to the amputated 1-point
function $\alpha$ from the symmetrized vertex number $1$. Using the Feynman
rules of section 3 one obtains

\begin{eqnarray}
\nonumber
\alpha_{[1]} (\eta) &=& 
	i \times i\kappa \times \frac12 
	\kappa \Hi \Om^{3-1/s} 
	\left[ \partial^{\alpha}(x) 
	\langle \psi(x) \psi_{0\alpha}(x') \rangle 
	\right]_{x' \rightarrow x} 
\\ 
\label{cont_v1}
 & = &  - \kappa^2 \Hi \Om^{3-1/s} \partial_\eta
	\left[
	- \frac{2}{s+1} i \Delta_A (x;x')
	- \frac{2s}{s+1} i \Delta_C (x;x')
	\right]_{x' \rightarrow x} \; .
\end{eqnarray}

We are looking for those terms that grow the fastest as a function of
time, such that after integration by (\ref{Inv_DA}) or (\ref{Inv_DC}) 
there can be a sustained effect of back reactions on the metric and
scalar field $A$, $C$ and $D$. Clearly, the vertex above fails to meet
that condition and contributes at maximum a constant to the expectation
values.

The terms which we are interested in arise when a time derivative hits
an external leg, as happens in vertex number 3, for example. When that
term is integrated by parts and the conformal time derivative hits the
volume factor $\Hi \Om^{3-1/s}$, the result is a factor of $\Hi^2
\Om^{4-2/s}$. This sort of term, when integrated with respect to the
inverse of the $A$-type propagator by (\ref{Inv_DA}), gives a
dominant logarithmic contribution.

On the other hand, when vertex number 3 is integrated with respect to the
inverse of the $C$-type propagator by (\ref{Inv_DC}), the result is only
a constant and other subdominant terms. Since integration by the $C$-type
propagator gives only subdominant contributions, we ignore them in what
follows.

It can be verified by inspection that no terms with factors that grow
faster than $\Om^{4-2/s}$ arise in the Einstein Lagrangian. Those are
the types of contributions we are looking for, and only the terms with
the right structure to contribute a factor of at least $\Hi^2
\Om^{4-2/s}$ will be collected in the following expressions.

Our results for the amputated 1-point functions to leading order are as
follows:

\begin{eqnarray}
\label{result_m}
\alpha(\eta) & = & - \frac12 \frac{1}{s+1} 
	\left( 3-\frac1s \right) \kappa^2 \Hi^2 \Om^{4-2/s}
	i \Delta_A^{(IR)} \; , \\
\label{result_n}
\gamma(\eta) & = & - \alpha(\eta) \; , \\
\label{result_u}
\delta(\eta) & = & - \frac{2}{\rs} \alpha(\eta) \; , \\
\end{eqnarray}
where $i\Delta_A^{(IR)}$ is given in expression (\ref{Prop_A_IR}).

The expectation values can be found with the help of expressions
(\ref{M})-(\ref{U}) and (\ref{Inv_DA}):

\begin{eqnarray}
\label{sol_M}
A(\eta) &=& 
	- \frac{\kappa^2 \Hi^2}{8\pi^2}
	\frac{s^2-s}{(s+1)^2} 
	\left[ 
	\frac{2 \Gamma \left(\frac32 + \frac{1}{s-1}\right)}{\sqrt{\pi}}
	\left( \frac{s-1}{s} \right)^{\frac{s}{s-1}}
	\right]^2 \\
\nonumber
&& \times \left[ \ln{\Om} - \frac{s}{3s-1} + 
	\frac{s}{3s-1}\Om^{-\frac{3s-1}{s}} \right] \; , \\
\label{sol_N}
C(\eta) &=& - \frac1s A(\eta) \; , \\
\label{sol_U}
\frac{D(\eta)}{\rs} &=& \frac1s A(\eta) \; ,
\end{eqnarray}
where we keep the subdominant terms after the logarithm only to
stress that expectation values vanish at the initial value surface
$\Om=1$. The prefactor in square brackets of expression
(\ref{sol_M}), which we call $\sigma(s)$, is approximately 1 for large
$s$.

The effective Hubble expansion parameter can be evaluated
by substituting the expressions above for the expectation value of
the metric into Eq. (\ref{H_eff}). It can be easily seen that
the logarithmic contributions cancel, and all that remains are
subdominant terms.
The effective expansion rate is
unchanged up to terms which are either constant and decaying as 
functions of time,

\begin{equation}
\label{H_eff_2}
{H}_{\rm eff}(\tilde{t}) = H(t) \left[ 1 + ({\rm subdominant}) \right]
\; .
\end{equation}

We conclude therefore that there is no one loop back reaction of 
quantum fluctuations on power law inflation to leading order.
Presumably there is a two loop
effect, analogous to the one found by Tsamis and Woodard \cite{Tsam}, 
that grows and would become dominant, but we did not attempt to calculate
those diagrams.

In the next section we will check the results above with a vastly
simpler canonical version of this calculation, and provide a physical
justification for the differences among the power-law and chaotic
inflation cases.

\section{The canonical result}

As a check on our previous results, and in its own right, it is
interesting to study the same problem of back reaction in power-law
inflation using canonical quantization. This was the method
used by Mukhanov {\it et al.} to derive their results.

In order to simplify the full Lagrangian (\ref{Lag}), we perform the
following approximations: first, ignore the spin-2 (gravity waves) and
spin-1 (``vector"  fluctuations) projections of the graviton degrees of
freedom, and concentrate on the scalar fluctuations of the metric that
couple to the fluctuations of the scalar field $\varphi$. Second, use the
constraints of the Einstein field equations explicitly in 
(\ref{Lag}) to eliminate the fluctuations of the scalar field and reduce
the number of degrees of freedom to one: the so-called Newtonian
potential $\Phi$. Third, the expectation values are given in terms of the
``spectrum" of the canonically quantized field $\Phi$ in power-law
inflation, which can be read right off the standard formulas in the
literature on quantum fluctuations in inflationary universe models
\cite{RevPaper,Lyth}.

The motivation for the first assumption lies in the observation that the
infrared limits of coincident propagators of the spin-1 and spin-2
degrees of freedom are dominated by the ultraviolet and fall off as a
function of time. Since cross-correlations are irrelevant at one loop, we
can ignore those degrees of freedom altogether and concentrate on the
dominant, scalar degrees of freedom of the graviton.

This truncated version of canonical quantization has obvious shortcomings, 
such as its
inadequacy to study perturbative corrections above lowest order and the
exclusion of many of the degrees of freedom from the calculation. These
difficulties would be unsurmountable if we wanted to to calculate quantum
corrections beyond one-loop, for example. Nevertheless, within the
scope of this one loop calculation the truncated canonical method is 
perfectly suited to study the leading contributions to back reaction.

The back reaction on the homogeneous and isotropic background is
determined by the expectation values of Einstein's field equations

\beq
\label{exp_EFE}
\langle 0 | G^{\mu}_{\nu} |0\rangle = 
	\frac{\kappa^2}{2} \langle 0| T^{\mu}_{\nu} |0\rangle
\eeq
and the equation of motion for the scalar field,
\beq
\label{EOM_corr}
\langle 0| \Box \varphi + V_{,\varphi} |0\rangle = 0 \; ,
\eeq
where the inflaton potential is

\beq
\label{V_fi_2}
V(\varphi) = 
	6 \left( \oots \right) \frac{\Hi^2}{\kappa^2} 
	\exp{ \left[ -\frac{\kappa}{\rs} \varphi \right] }\; .
\eeq

The terms in Einstein's equations which are quadratic in the quantum
fluctuations can be collected in an {\it effective} stress-energy tensor
$\tau_{\mu\nu}$, which is a source term in addition to the stress-energy
tensor of the background matter \cite{Mukh,Abra}. The equation of motion
for $\varphi$ is similarly corrected by source terms quadratic in the
quantum fields. In this section we shall regard back reaction as the
response of the background to the source terms induced by quantum
fluctuations, found by solving for the expectation values of Einstein's
equations and the equation of motion for the scalar field.

We write the metric in longitudinal gauge,

\beq
\label{ds2}
ds^2 =	- \left[ 1 + 2\Phi(t,\vec{x}) \right] dt^2  
	+ a_0^2 (t) \left[ 1 - 2\Phi(t,\vec{x}) + 2w(t)\right] 
	d\vec{x} \cdot d\vec{x} \; ,
\eeq
and the scalar field is

\beq
\label{scalar}
\varphi = \varphi_0(t) + \phi(t,\vec{x}) + v(t) \; ,
\eeq
where $\Phi$ and $\phi$ are, as before, quantum fields which ought to be
canonically quantized, and $w(t)$ and $v(t)$ are respectively corrections
to the scale factor and to the scalar field from back reaction\footnote{We have 
adopted a notation different from earlier sections to avoid confusion, since
we in the section we work in a different gauge}. We
have taken advantage of the freedom of gauge at the second order to fix any
correction to the time slicing (the $N$ of earlier sections) to zero. 
Therefore, in this gauge $\dot{w}$ is the correction to the Hubble
expansion rate, $H_{\rm eff} = H + \dot{w}$ that will eventually be
compared with the effective expansion rate (\ref{H_eff_2}) of the last
section.

The spectrum $|\delta_\Phi(k,t)|$ is defined by the expectation value of
the square of the canonically quantized Newtonian potential: 

\beq
\label{Exp_Phi}
\langle 0| \Phi (t,\vec{x}) \Phi(t,\vec{x}) |0\rangle =
\int \frac{dk}{k} \left| \delta_\Phi (k,t) \right|^2 \; ,
\eeq
The spectrum is proportional to $k$ when $k \rightarrow \infty$, and
consequently the expectation value above diverges as $k^2$ in the far
ultraviolet. Just as was done in section 5, we assume that a proper
renormalization of these infinities have been performed and that the
appropriate counterterms have no bearing on the long-range
interactions described by the effective theory.

The expectation value (\ref{Exp_Phi}) is also divergent in the infrared
since the spectrum is approximately constant for very small momenta.
Again the solution is to work in a compact spatial
manifold, where $k$ is cut off at the
value $\Hi$ corresponding to the radius of the manifold at the initial
time $t=0$.

The infrared limit of the spectrum of the Newtonian potential during
power-law inflation is \cite{Lyth}

\beq
\label{delta_Phi}
|\delta^{(IR)}_\Phi (k,t)|^2 = 
	\frac{\kappa^2 \Hi^2}{32\pi^2}
	\frac{\sigma(s)}{s} 
	\left( \frac{k}{\Hi} \right)^{-\frac{2}{s-1}} \; ,
\eeq
where 

\beq
\label{sigma_s}
\sigma(s) = \left[ 
	\frac{2	\Gamma \left( \frac32 + \frac{1}{s-1} \right) }{\sqrt{\pi}} 
	\left( \frac{s-1}{s} \right)^{\frac{s}{s-1}}
	\right]^2 \; ,
\eeq 
is the same factor defined below (\ref{sol_M}), and approaches 1 as $s
\gg 1$.

The physical information contained in this spectrum is that the
amplitudes of long-wavelength fluctuations are asymptotically constant in
power-law inflation. The amplitudes of infrared fluctuations in chaotic
inflation, by comparison, grow slowly with time. Where back reaction
is concerned, quantum corrections to power-law inflation from a
fluctuation with a fixed comoving wavelength can be at most constant,
while corrections to the background in chaotic models can grow as a
function of time during the inflation of the universe.

The distinguishing facts about power-law inflation are 1) the equation of
state $w\equiv\rho/p=-1+2/3s$ is constant throughout inflation, and 2) 
the kinetic and potential energy densities of the background scalar
field, as well as the Hubble parameter, are at fixed ratios with respect
to each other at any given time. Consequently, the amplitude of
fluctuations on large scales, which couple to these ratios, freeze to a
constant value.

We often find claims in the literature to the effect that
quantum fluctuations during inflation freeze after they become larger
than the Hubble radius. While this is exact for power-law inflation, it
is only approximate for most models of inflation.

With the spectra above, the expectation value of the Newtonian potential
in the infrared limit is, by (\ref{Exp_Phi}),

\beq
\label{exp_NP}
\langle 0 | \Phi(t,\vec{x}) \Phi(t,\vec{x}) |0\rangle =
	\sigma(s) \frac{s-1}{s} 
	\frac{\kappa^2 \Hi^2}{32 \pi^2} = {\rm const} \;
.
\eeq
Notice that this is in accord with expressions (\ref{Prop_metric}) and
(\ref{Prop_A_IR}) for large $s$ (remember that $\psi_{00}=2\Phi$).
The expectation value of the scalar field can be deduced from this
expression by using the following useful constraint in momentum space,
valid in longitudinal gauge:

\beq
\label{id_scf}
\phi(t,k) = - 2 \frac{V}{V_{,\varphi}} \Phi(t,k) 
	  = 2 \frac{\rs}{\kappa} \Phi(t,k) \; .
\eeq

The Einstein equations with quantum corrections are found by using the
metric (\ref{ds2}) and scalar field (\ref{scalar}) into Eqs. 
(\ref{exp_EFE}). The {\it general} result (the spatial gradient terms
have been ignored since we consider only infrared modes) is: 

\begin{eqnarray}
\label{tauzero}
3H^2 & + & 3\langle \dot{\Phi}^2 \rangle 
	 + 12 H^2 \langle \Phi^2 \rangle 
	+ 6 H \dot{w} 
	= \frac{\kappa^2}{2} \left[ \frac12 \dot\varphi_0^2 + V \right.
\\ \nonumber
& + & 	\left. \frac12 \left( 
	\langle \dot{\phi}^2 \rangle 
	- 4 \dot\varphi_0 \langle \dot\phi \Phi \rangle
	+ 4 \dot\varphi_0^2 \langle \Phi^2 \rangle
	+ 2 \dot\varphi_0 \dot{v} \right) 
	+ \frac12 V_{,\varphi \varphi} \langle \phi^2 \rangle
	+ V_{,\varphi} v \right] \; ,
\end{eqnarray}

\begin{eqnarray}
\label{tauij}
3H^2 & + & 2\dot{H} 
	+ 4 \left( H^2+2\dot{H} \right) \langle \Phi^2 \rangle 
	+ 8 H \langle \dot\Phi \Phi \rangle
	+ \langle \dot\Phi^2 \rangle
	+ 6 H \dot{w}
	+ 2 \ddot{w}  
\\ \nonumber
& = &  \frac{\kappa^2}{2} \left[ 
	- \frac12 \dot\varphi_0^2 + V  
	- \frac12 \left( 
	\langle \dot{\phi}^2 \rangle 
	- 4 \dot\varphi_0 \langle \dot\phi \Phi \rangle
	+ 4 \dot\varphi_0^2 \langle \Phi^2 \rangle
	+ 2 \dot\varphi_0 \dot{v} \right) \right.
\\ \nonumber
& &   	+ \left. \frac12 V_{,\varphi \varphi} \langle \phi^2 \rangle
	+ V_{,\varphi} v \right] \; ,
\end{eqnarray}
where brackets $\langle \cdots \rangle$ denote vacuum expectation values.

As we have argued in the discussion of the spectrum, terms containing
time derivatives of $\Phi$ and $\phi$ should vanish. Using the identity
(\ref{id_scf}), we can finally write the Einstein equations with
quantum corrections above in terms of only $\langle \Phi^2 \rangle $,
$w(t)$ and $v(t)$:

\begin{eqnarray}
\label{efe_2_0}
6H\dot{w} & = & \frac{\kappa^2}{2} 
	\left[	
	\dot\varphi_0 \dot{v} 
	- V \left( 
	  \frac{\kappa v}{\rs} 
	  + 2 \langle \Phi^2 \rangle \right) \right] \; , 
\\ \label{efe_2_i}
6H\dot{w} + 2 \ddot{w} & = & \frac{\kappa^2}{2} 
	\left[	
	\dot\varphi_0 \dot{v} 
	- V \left( 
	  \frac{\kappa v}{\rs} 
	  + 2 \langle \Phi^2 \rangle \right) \right] \; . 
\end{eqnarray}

It is also instructive to write down the equation of motion for the
scalar field to second order,

\begin{eqnarray}
\label{eom_2}
\ddot{v} &+& 3H\dot{v} 
	+ 3\dot{w} \dot\varphi_0 
	+ V_{,\varphi\varphi}v \\ \nonumber
	&-& 4 \langle \dot\Phi \dot\phi \rangle 
	- 4 \dot\varphi_0 \langle \dot\Phi \Phi \rangle 
	+ 2 V_{,\varphi\varphi} \langle \Phi \phi \rangle 
	+ \frac{1}{2} V_{,\varphi\varphi\varphi} \langle \phi^2 \rangle 
	= 0 \; ,
\end{eqnarray}
which, after use of (\ref{id_scf}), the background
identities and the Einstein field equation (\ref{efe_2_0})
read

\begin{eqnarray}
\label{eom_2_i}
t^2 \ddot{x} + \frac{3 s + 1 }{s} t \dot{x} + 
	3 s \left( x + 2 \langle \Phi^2 \rangle \right) = 0 \; ,
\end{eqnarray}
where $x=\kappa v/\rs$. Since this equation is second order in time, we
can always find solutions such that $v(t_0)=\dot{v}(t_0)=0$. Notice that
$\langle \Phi^2 \rangle$ is a constant by (\ref{exp_NP}).
It is straightforward to solve this equation, and the result is

\beq
\label{sol_x}
x = \frac{\kappa v}{\rs} =  
	-2 \left[ 1 
	- \frac{3s-1}{3s-2} \frac{t_0}{t} 
	+ \frac{1}{3s-2} \left( \frac{t_0}{t} \right)^{3s-1} \right] 
	\langle \Phi^2 \rangle \; ,
\eeq
where we have chosen the integration constants such that
at the initial value surface $x(0)=\dot{x}(0)=0$.

The dominant contribution to $x=\kappa v /\rs$ is therefore a constant,
corresponding to the homogeneous solution to (\ref{eom_2_i}). However, as
happened in the covariant calculation, we have neglected subdominant
terms which decay as a function of time in the present calculation, by
discarding several of the degrees of freedom of the fundamental
Lagrangian. Therefore, the inhomogeneous solutions are unreliable within
our approximation scheme and will be discarded.

Substituting the homogeneous part of (\ref{sol_x}) into (\ref{efe_2_i})
we get the following equation for the correction to the scale factor:

\beq
\label{eq_w_x}
	t^2 \ddot{w} + 3 s t \dot{w} = 0 \; .
\eeq
The solution is also straightforward, and we get

\beq
\label{sol_w}
\dot{w} = 0 \; 
\eeq
exactly since the decaying solution $\dot{w} \propto t^{-3s}$ has to be
zero to satisfy the retarded boundary conditions.

The effective expansion rate is therefore given by,

\begin{eqnarray}
\label{H_effective}
H_{\rm eff} &=& H(t) \; .
\end{eqnarray}

We have thus obtained the back reaction on the metric and
scalar field by finding the solutions to the expectation values of the
Einstein field equations (\ref{exp_EFE}) and the equation of motion for
the scalar field (\ref{EOM_corr}). Expectation values were evaluated
using the canonically quantized Newtonian potential $\Phi$.

The result of the canonical calculation is that the back reaction of
quantum fluctuations during power-law inflation does not affect the
expansion rate of the universe, at least to leading order. This is in
agreement with the results of the covariant calculation of last section.

As pointed out earlier in this section, infrared fluctuations of a fixed
comoving wavelength in power-law inflation have constant amplitudes. Those
fluctuations exiting the Hubble radius at a later time in the inflation
epoch have a smaller amplitude than the ones that exited earlier, since
the scale of inflation is decreasing as $\propto t^{-2}$. The cumulative
effect of superimposing modes of different comoving momenta is not
sufficient to make the expectation values of the quantum fields grow in
time, and the momentum mode sum is dominated by modes that exited the
horizon early in the inflation epoch.

\section{Discussion}

We have calculated the back reaction of quantum fluctuations on the
expansion rate of homogeneous backgrounds of power-law inflation models.
Two methods were employed: covariant quantization of the full
scalar-graviton system, and canonical quantization of a reduced system
where the spin-1 and spin-2 degrees of freedom of the graviton were
purged.

The results of the two calculations are identical: to leading order, there
is no effect of the quantum fluctuations on the effective Hubble parameter
in power-law inflation. The physical reason is that long wavelength modes
have constant amplitudes, and therefore both the amputated 1-point
functions and the source terms in the expectation values of Einstein's
equations are constant.

The main result is that the shape of the inflaton potential can have an
enormous impact on back reaction. While chaotic inflation can have a
significant back reaction, power-law inflation does not.

We also hope that the results presented here (see also \cite{Wood}) will
help settle some of the criticisms raised with respect to the canonical
method as applied to the study of back reaction in quantum general
relativity.

\vskip 1cm

\centerline{\bf Acknowledgments}

It is a pleasure to acknowledge stimulating and informative conversations with 
R. H. Brandenberger and V. F. Mukhanov. We are also grateful to the University 
of Crete for its hospitality at the inception of this project. This work was 
partially supported by DOE contract DE-FG02-97ER\-41029, by NSF grant 
94092715, by NATO grant CRG-971166 and by the Institute for Fundamental Theory.

\end{document}